\documentclass[a4paper,10pt,twoside]{cpc-hepnp}

\usepackage{multicol}
\usepackage{graphicx}
\usepackage{booktabs}
\usepackage{amssymb,bm,mathrsfs,bbm,amscd}
\usepackage[tbtags]{amsmath}
\usepackage{lastpage}
\usepackage{CJK}

\begin{document}
\begin{CJK*}{GB}{gbsn}

\fancyhead[c]{\small Li-Zhu Chen~ et al: The sixth order cumulant of  net-proton number in Binomial distribution  at  $\sqrt{s_{NN}} = $ 200 GeV}


\title{The sixth order cumulant of net-proton number in Binomial distribution  at  $\sqrt{s_{NN}} = $ 200 GeV\thanks{This work was supported by the postdoctral science and technology project of Hubei Province under Grant No. 2018Z27, and the Fundamental Research Funds for the Central Universities under Grant No. CCNU19ZN019.}}

\author{%
Li-Zhu Chen$^{1;1)}$\email{chenlz@nuist.edu.cn}%
\quad Ye-Yin Zhao$^{2}$
\quad Jin Wu $^{2}$
\quad Zhi-Ming Li$^{2}$
\quad Yuan-Fang Wu$^{2}$
}
\maketitle

\address{%
$^1$School of Physics and Optoelectronic Engineering, Nanjing University of Information Science and Technology, Nanjing 210044, China\\
$^2$Key Laboratory of Quark and Lepton Physics (MOE) and
Institute of Particle Physics, Central China Normal University, Wuhan 430079, China\\
}

\begin{abstract}
It is proposed that ratios of the sixth order to the second order cumulant ($C_6/C_2$) of conserved quantities are sensitive to the chiral crossover transition.  Recently, the negative $C_6/C_2$ was obtained both in theoretical Lattice QCD and experiments at   $\sqrt{s_{NN}} = $ 200 GeV. In this study,
we investigate the  behavior of net-proton $C_6/C_2$ in statistical Binomial distribution (BD) at  $\sqrt{s_{NN}} = $ 200 GeV in Au + Au collisions.
With the  BD parameters extracted from RHIC/STAR, it is found that  $C_6/C_2$ can be negative.  Furthermore,  the obtained $C_6/C_2$ becomes smaller
when applying the same magnitude of experimental statistics and calculation method to simulations.  In 0-10\% centrality, there is a significant difference between the simulated result and theoretical expectation.   Based on the extracted parameters and experimentally collected statistics, the baseline of net-proton $C_6/C_2$ in BD is presented.
\end{abstract}



\begin{multicols}{2}

\section{Introduction}
One of the main goals of the Beam Energy Scan (BES) Program at the Relativistic Heavy Ion Collider (RHIC) is to explore the phase diagram of the Quantum Chromodynamics (QCD)~\cite{STAR-BESI}. At vanishing and small values of the chemical potentials for conserved charges, such as net-baryon, net-charge and net-strangeness, Lattice QCD has predicted that it occurs a smooth crossover from the hadronic phase to the QGP phase~\cite{crossover}.   Based on Lattice QCD calculations with physical values of light and strange quark masses,  the negative signal of ratio of the sixth order to the second order ($C_6/C_2$) is observed in the crossover region~\cite{C6-karsch-v1,C6-karsch-v2,C6-karsch-v3,C6-karsch-v4}.  The QCD-assisted low-energy effective theory and the QCD based models, such as the Polyakov loop extended quark-meson (PQM) and the Nambu-Jona-Lasinio (PNJL) models, also support that the sixth order cumulant is negative near the chiral crossover transition~\cite{PQM-C6-v0,PQM-C6-v1,PQM-C6-v2, PNJL-C6-v1, PNJL-C6-v2}. Consequently, $C_6/C_2$ is a powerful observable to study the QCD phase diagram in experiment.

The STAR Collaboration had reported the preliminary results of  net-proton $C_6/C_2$~\cite{STAR-C6C2-v1,STAR-C6C2-v2, STAR-C6C2-v3}.  In 0-40\% centrality,  $C_6/C_2$ is negative at $\sqrt{s_{NN}}$ = 200 GeV, which is consistent with the Lattice QCD calculation.  However,   $C_6/C_2$ is positive and close to unity at $\sqrt{s_{NN}}$ = 54.4 GeV, unlike the QCD predicted negative behavior. In the future, with the sufficient statistics accumulated, the analysis of the net-proton $C_6/C_2$ is possible by the ALICE collaboration at LHC~\cite{LHC}.
Before illustrating the physics of measured $C_6/C_2$, the contributions of non-phase transition related influences should be subtracted, such as  the conservation of the total baryon number, experimental acceptances in terms of kinematic variables,  efficiency corrections, the difference of $C_6/C_2$ in net-proton and net-baryon  and so on~\cite{C6C2-koch, C6C2-influence-1, C6C2-influence-2, C6C2-influence-3}. It was shown that $C_6/C_2$ can be negative for $\sqrt{s_{NN}} \le $ 40 GeV due to the baryon number conservation~\cite{C6-fu}. In this paper, we will focus on discussions  of the BD contribution and influence of the statistics used for data  analysis at  $\sqrt{s_{NN}}$ = 200 GeV.

In nuclear collisions, some basic statistical distributions, such as the Poisson distribution, BD and Negative Binomial distribution (NBD), are frequently used to describe the shape of the multiplicity distributions~\cite{HRG-1,HRG-2,bd-1,bd-2}.
 Theoretically, in the free hadron gas in equilibrium, the resonance gas in the hadron phase obeys the Poisson distributions. It was also argued that the proton number is given by the superposition of the binomial distribution of the baryon number due to the isospin randomization~\cite{Asakawa}. Consequently, studies of the cumulants in different statistical distributions can extract information related to the nature of the particle production mechanism.

Recently, cumulants in these statistical distributions were widely taken as baselines to help us understand the experimentally measured cumulants, such the the cumulants of the net-proton, net-charge, net-kaon and net-$\Lambda$~\cite{NBD-gary,Xiaofeng-baseline-proton, STAR-proton-v1, STAR-charge,PHENIX-charge,STAR-kaon, STAR-kappa}. To date, the studies of cumulants in BD/NBD have been up to the fourth order. It is found that the BD/NBD baselines can quantitatively explain the cumulants of net-charge, net-kaon and net-$\Lambda$ in experiment. The BD can also describe the fourth order cumulant of the net-proton number  at $\sqrt{s_{NN}}\ge$ 39 GeV.  Consequently, it is also interesting to study the net-proton $C_6/C_2$ in BD.

With the data accumulated, STAR has about 850M events for analysis of $C_6/C_2$ at  $\sqrt{s_{NN}}$ = 200 GeV~\cite{STAR-C6C2-v2}.  However,  the data used in the analysis include three independent parts: (a) around 420M events for 0-80\% centrality collected from the year 2010 minimum bias (MB) trigger, (b) around 110M  events for 0-10\% centrality from the year 2010 central trigger, and (c) around 320M events for 0-80\% centrality from the year 2011 MB trigger. In principle, $C_6/C_2$ in each centrality should be firstly calculated by these three independent parts separately. For each part, $C_6/C_2$ is obtained by the method of centrality bin width correction (CBWC)~\cite{cbwc-star,cbwc-v2,cbwc-v3}. It means  $C_6/C_2$ needs to be calculated  first in each Refmult3 bin ($N_{ch}$) to reduce the initial size fluctuation. Then, $C_6/C_2$ is averaged over all $N_{ch}$ in a given centrality.

 A critical issue is that the statistics in each of $N_{ch}$ are still considerably limited even with 320M MB events.  Our previous studies showed that we must check whether the statistics in each of $N_{ch}$ are sufficient to satisfy the Central Limit Theorem (CLT) when applying the CBWC method in cumulants analysis~\cite{statistics-chenlz-v1,statistics-chenlz-v2,statistics-chenlz-v3}.
It showed that 90M MB events are not sufficient for analysis of net-proton $C_6/C_2$  at $\sqrt{s_{NN}} = 11.5$ GeV in the UrQMD model.  These statistics are of the same order of magnitude as that in each data collection at $\sqrt{s_{NN}} = $ 200 GeV at RHIC/STAR.
In addition,  the required statistics are also related to the detail shape and width of the net-proton multiplicity distributions.  Therefore,  we need to re-examine if we can obtain reliable $C_6/C_2$  with 320M MB events.

\begin {figure*}[htbp]
\begin{center}
\includegraphics[width=2.73in]{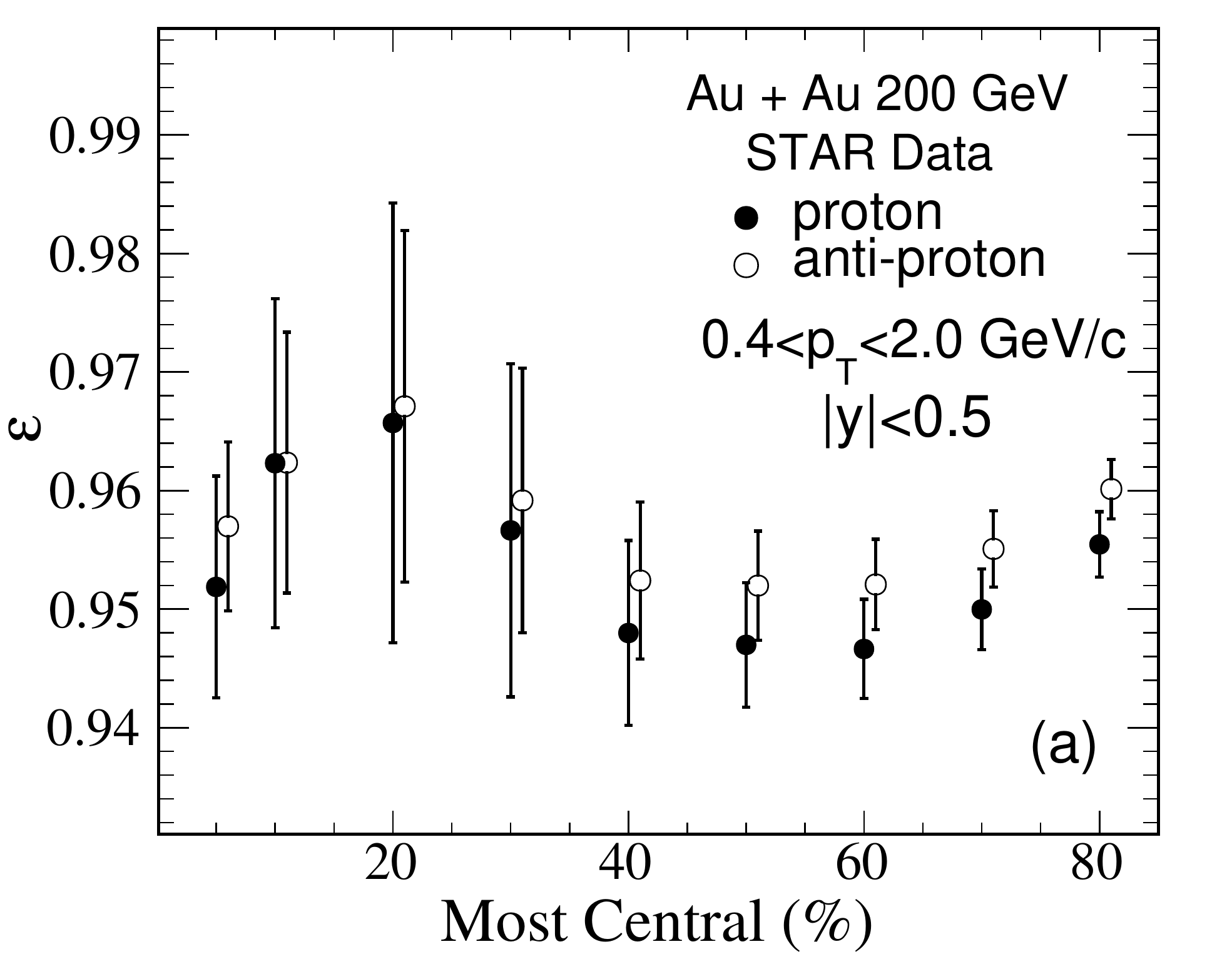}
\includegraphics[width=2.73in]{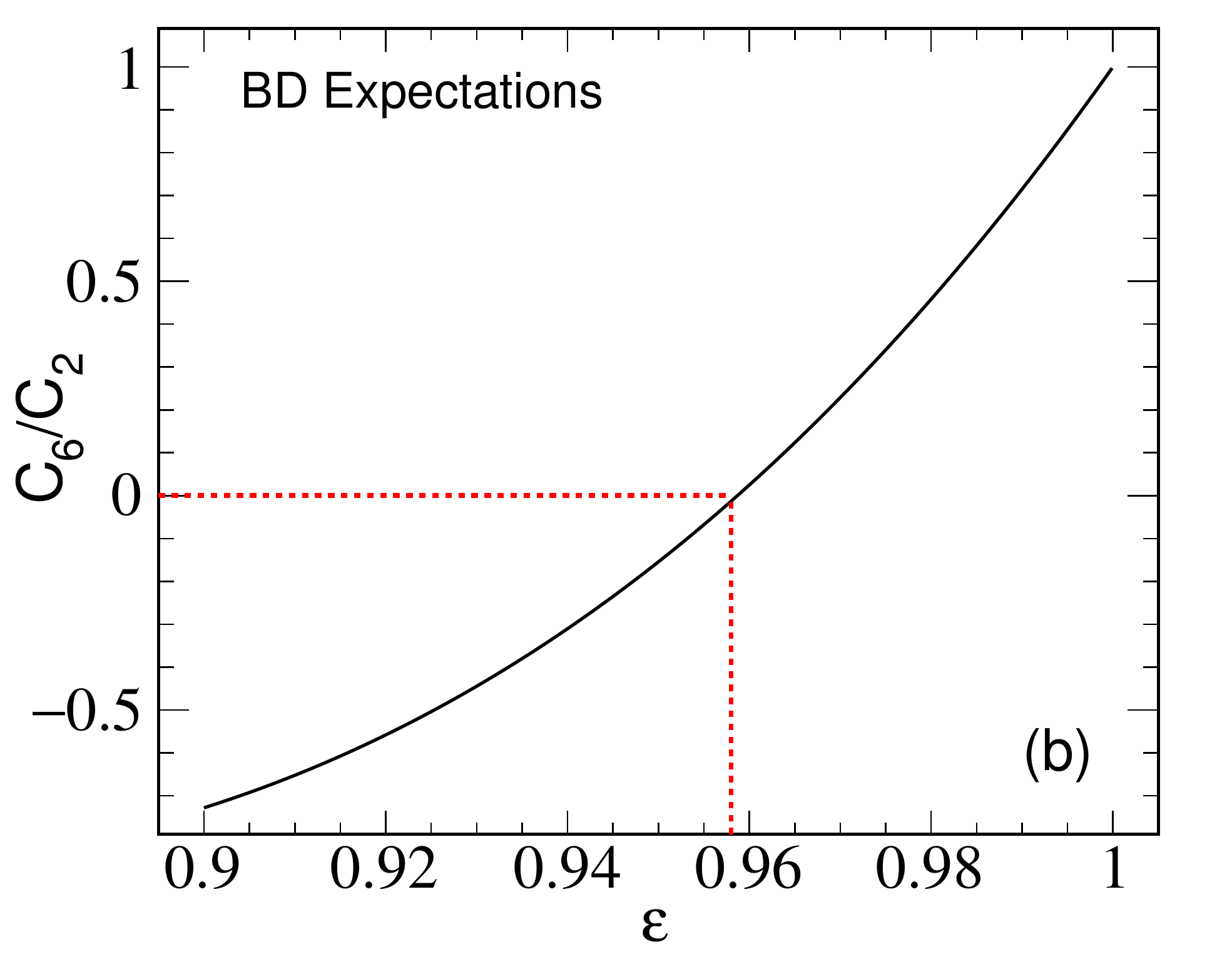}
\figcaption{\label{C6-bd-expectations} (Color Online) The left panel shows $\varepsilon_p$  and $\varepsilon_{\bar{p}}$ in different centralities  at $\sqrt{s_{NN}} =$ 200 GeV in Au + Au collisions measured by RHIC/STAR~\cite{cbwc-v3}. The protons and anti-protons are identified at mid-rapidity $(|y|<0.5)$ within  $0.4<p_{T}< 2.0$ GeV/$c$. The right panel is the  $\varepsilon$ dependence of   $C_6/C_2$.}
\end{center}
\end{figure*}

In this paper, we will start off the discussions from  behavior of net-proton  $C_6/C_2$ in BD.
 At $\sqrt{s_{NN}}$ = 200 GeV, it is found that the negative $C_6/C_2$ can be obtained with the experimentally measured parameters.
To study influence of the statistics on $C_6/C_2$, a toy simulation about validation of the CLT in $C_6/C_2$ is demonstrated in Section 3. To make a direct comparison with experimental data, more detail procedures of simulations are described
 in Section 4.
With the experimentally collected statistics and extracted BD parameters, we show that there is a marked drop in $C_6/C_2$ in 0-10\% centrality in Section 5.   The baseline of  $C_6/C_2$ is also demonstrated.  Finally, the summary is presented in Section 6.

\section{$C_6/C_2$ in BD}
In probability theory and statistics, the BD is the discrete probability distribution of the number of successes in a sequence of $n$ independent binary experiments, where the results of each experiment is true with probability $p$ and false with probability $q=1-p$~\cite{bimomial-distribution}. The probability that the binomial random variable $x$ takes on values in its range can be expressed using the binomial probability function:

\begin{equation}\label{bd-formula}
P(x) = \left(\begin{array}{c}
n\\x \end{array} \right)p^{x}(1-p)^{n-x}=\frac{n!}{x!(n-x)!}p^{x}(1-p)^{n-x},
\end{equation}
where $x$ corresponds to the number of protons or anti-protons in each event.

Experimentally, if we know the mean $\mu$ and variance $\sigma^2$ ($\sigma^2<\mu$ for a BD, while $\sigma^2>\mu$ for a NBD) of the distribution, then the input parameters of $p$ and $n$ are:

\begin{equation}\label{bd-probability}
p=1-\frac{\sigma^2}{\mu}=1-\varepsilon,
\end{equation}

and

\begin{equation}\label{bd-probability}
n=\frac{\mu}{p}=\frac{\mu}{1-\varepsilon},
\end{equation}
where  $\varepsilon=\frac{\sigma^2}{\mu}$.

With given $\mu$ and $\varepsilon$, the expectations of cumulants from the second to the sixth order can be written as:
\begin{eqnarray}
C_2 &=& n\left(p-p^2\right) = \varepsilon\mu\label{2nd-exp}\\
C_3 &=& n\left(p-3p^2+2p^3\right) = \varepsilon\mu \left(-1+2\varepsilon\right)\label{3rd-exp}\\
C_4 &=& n\left(p-7p^2+12p^3-6p^4\right)= \varepsilon\mu \left(1-6\varepsilon + 6\varepsilon^2\right)\label{4th-exp}\\
C_5 &=& n\left(p-15p^2+50p^3-60p^4+24p^5\right)\nonumber\\ &=& \varepsilon\mu \left(-1 + 14\varepsilon - 36 \varepsilon^2 + 24 \varepsilon^3\right)\label{5th-exp}\\
C_6  &=&  n\left(p-31p^2+180p^3-390p^4+360p^5-120p^6\right)\nonumber\\ & =&  \varepsilon\mu \left(1-30\varepsilon + 150\varepsilon^2 - 240\varepsilon^3 + 120 \varepsilon^4\right)\label{6th-exp}
\end{eqnarray}

If the numbers of protons and anti-protons are independently produced as BD,  the net-proton $C_6/C_2$  can be expressed as:
\begin{equation}\label{C6C2-bd}
C_6/C_2 = \frac{C_6^{p}+C_6^{\bar{p}}}{C_{2}^{p}+C_{2}^{\bar{p}}}.
\end{equation}

Generally, the expected $C_6/C_2$ is related to $\varepsilon_p$,  $\varepsilon_{\bar{p}}$, $\mu_p$, and $\mu_{\bar{p}}$.  Based on these four parameters, one can obtain the expected $C_6/C_2$ in each centrality.  In contrast, the experimental studies had shown that $\varepsilon_p$ and $\varepsilon_{\bar{p}}$ are close to each other  at $\sqrt{s_{NN}} = 200$ GeV in Au + Au collisions  shown in   Fig.~\ref{C6-bd-expectations}(a) ~\cite{cbwc-v3}.   The error contains the statistical and systematical uncertainties, which is performed by $\sigma_{\varepsilon} =\sqrt{\sigma^2_{stat} + \sigma^2_{sys}}$. The protons and anti-protons are selected at mid-rapidity $(|y|<0.5)$ within  $0.4<p_{T}< 2.0$ GeV/c.
It shows $\varepsilon_p$ and $\varepsilon_{\bar{p}}$ extracted from STAR are consistent with each other.  Within  1$\sigma_{\varepsilon}$ of uncertainty, the centrality dependence of $\varepsilon$ is weak.
To  make an appropriate approximation, we assume $\varepsilon = \varepsilon_p = \varepsilon_{\bar{p}}$ in this paper. In this case,
the expectation of net-proton $C_6/C_2$ can be written as:
\begin{equation}\label{C6C2-bd}
C_6/C_2 = 1-30\varepsilon + 150\varepsilon^2 - 240\varepsilon^3 + 120 \varepsilon^4.
\end{equation}

Eq.~(\ref{C6C2-bd}) shows the expectation of  $C_6/C_2$ is only dependent on $\varepsilon$.  The effects of  $\mu_p$ and $\mu_{\bar{p}}$ are cancelled. The detailed $\varepsilon$ dependence of $C_6/C_2$ is shown in  Fig.~\ref{C6-bd-expectations}(b).
It shows $C_6/C_2$ drops drastically with the decrease of $\varepsilon$.
It is already negative with $\varepsilon<0.958$, which is within the experimentally measured range.
 In addition,  $C_6/C_2$ has a broad value range and it can change from positive to negative,
within 1$\sigma_{\varepsilon}$ uncertainty of $\varepsilon$.
Consequently, it is not suitable to directly give the expectation only based on the unique measured value of $\varepsilon$, without considering its uncertainty. Instead, we must set an interval of $\varepsilon$ to study the behavior of $C_6/C_2$.     Fig.~\ref{C6-bd-expectations}(a) shows the upper and lower values of the error bar touch $\varepsilon$ at approximately 0.99 and 0.94, respectively. If we assume the range of $\varepsilon$ is between 0.94 and 0.99, then the expected values of  $C_6/C_2$ are from  -0.31 to 0.71. The negative $C_6/C_2$ can be obtained in the pure statistical BD. Only the negative signal is not enough to be taken as an indication of a smooth crossover transition.

Here the obtained $C_6/C_2$ is the ideal theoretical expectation. In the experiment, the statistics are still a critical issue for the analysis of $C_6/C_2$. The satisfaction of the CLT requires to be carefully checked before the data analysis.

\section{Central Limit Theorem in $C_6/C_2$}
\begin {figure*}
\begin{center}
\includegraphics[width=2.7in]{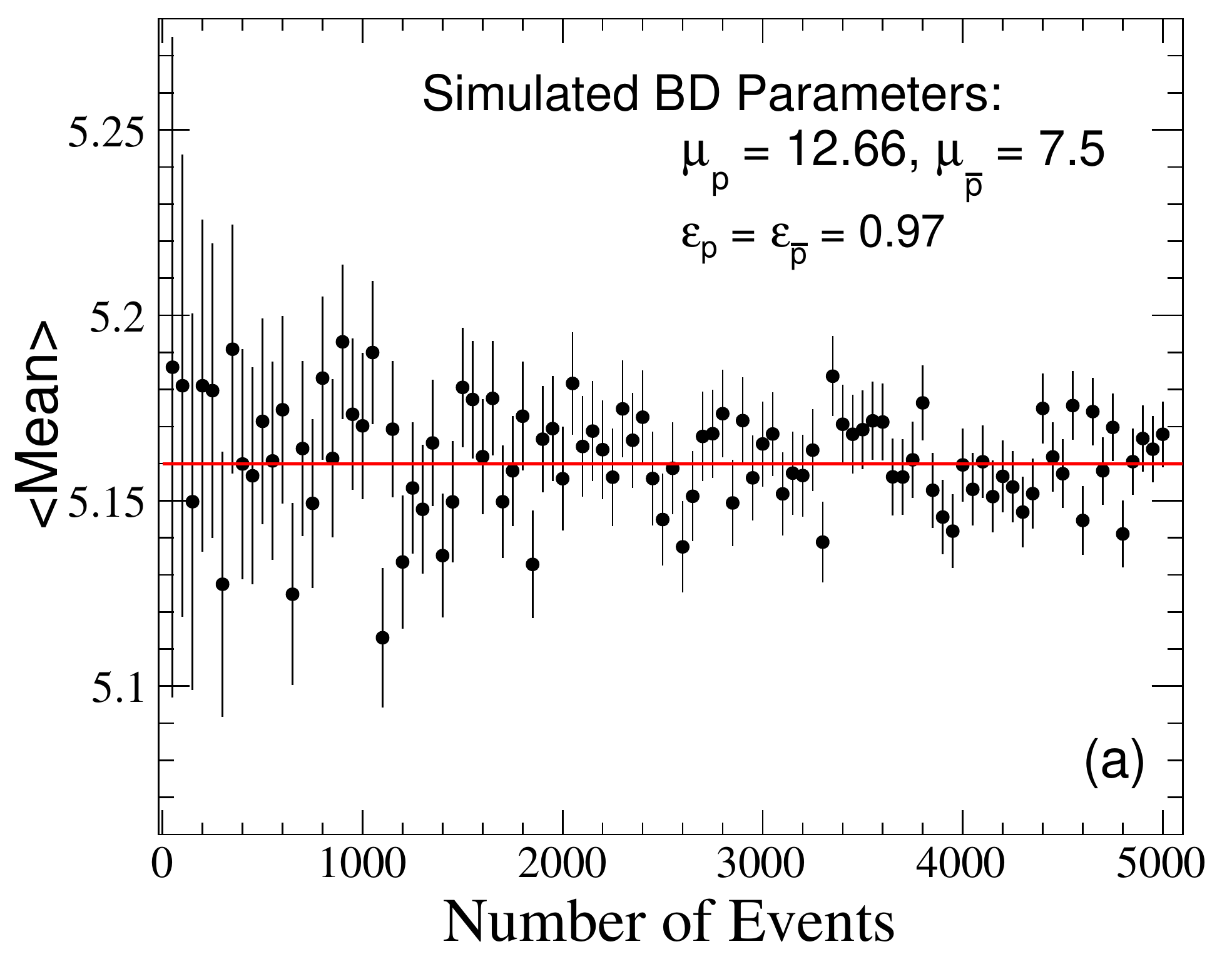}
\includegraphics[width=2.7in]{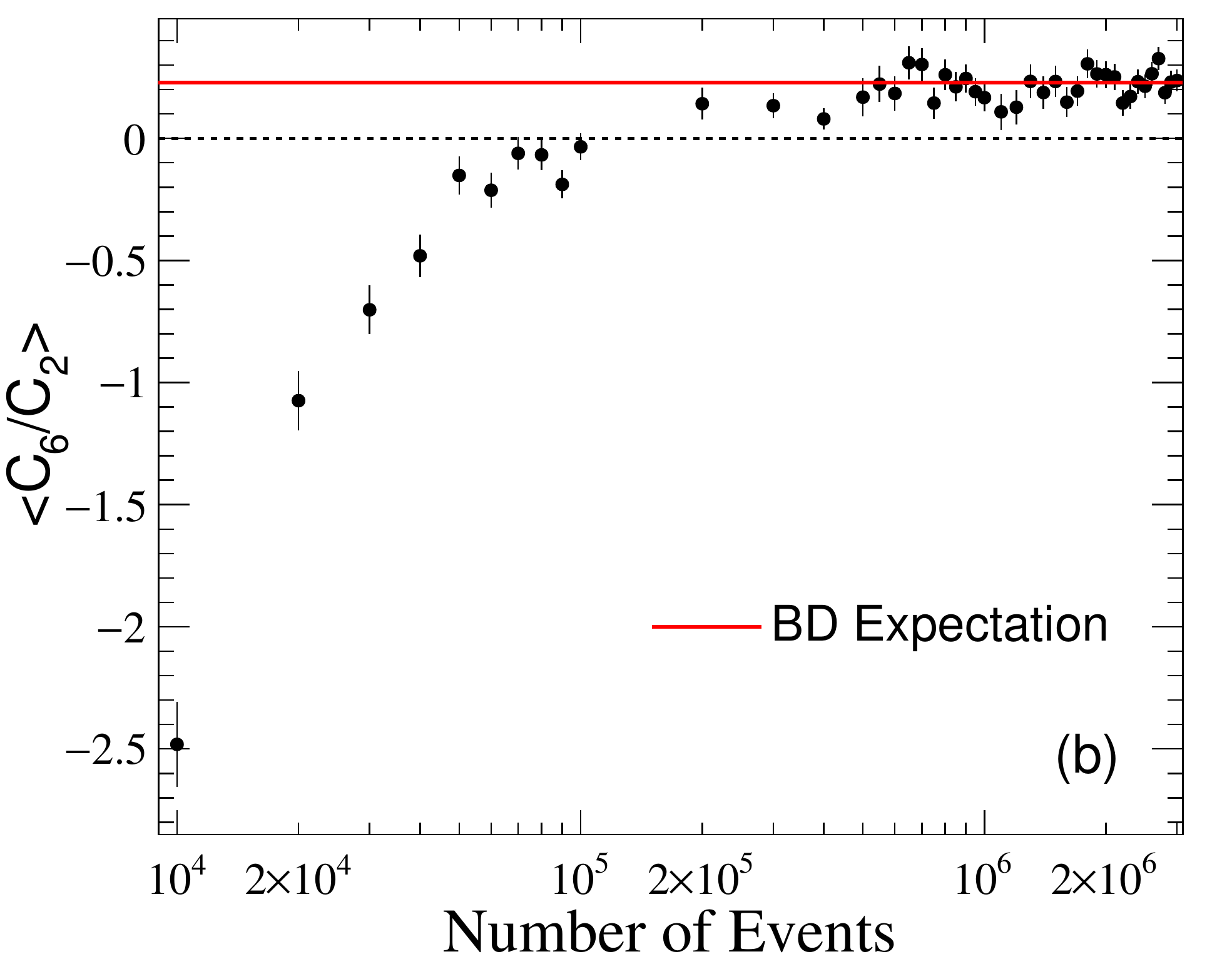}
\figcaption{\label{mean-clt} (Color Online) Statistics dependence of $\left<Mean\right>$ and $\left<C_6/C_2\right>$, respectively.}
\end{center}
\end{figure*}

\begin{figure*}[b]
\begin{center}
\includegraphics[width=2.1in]{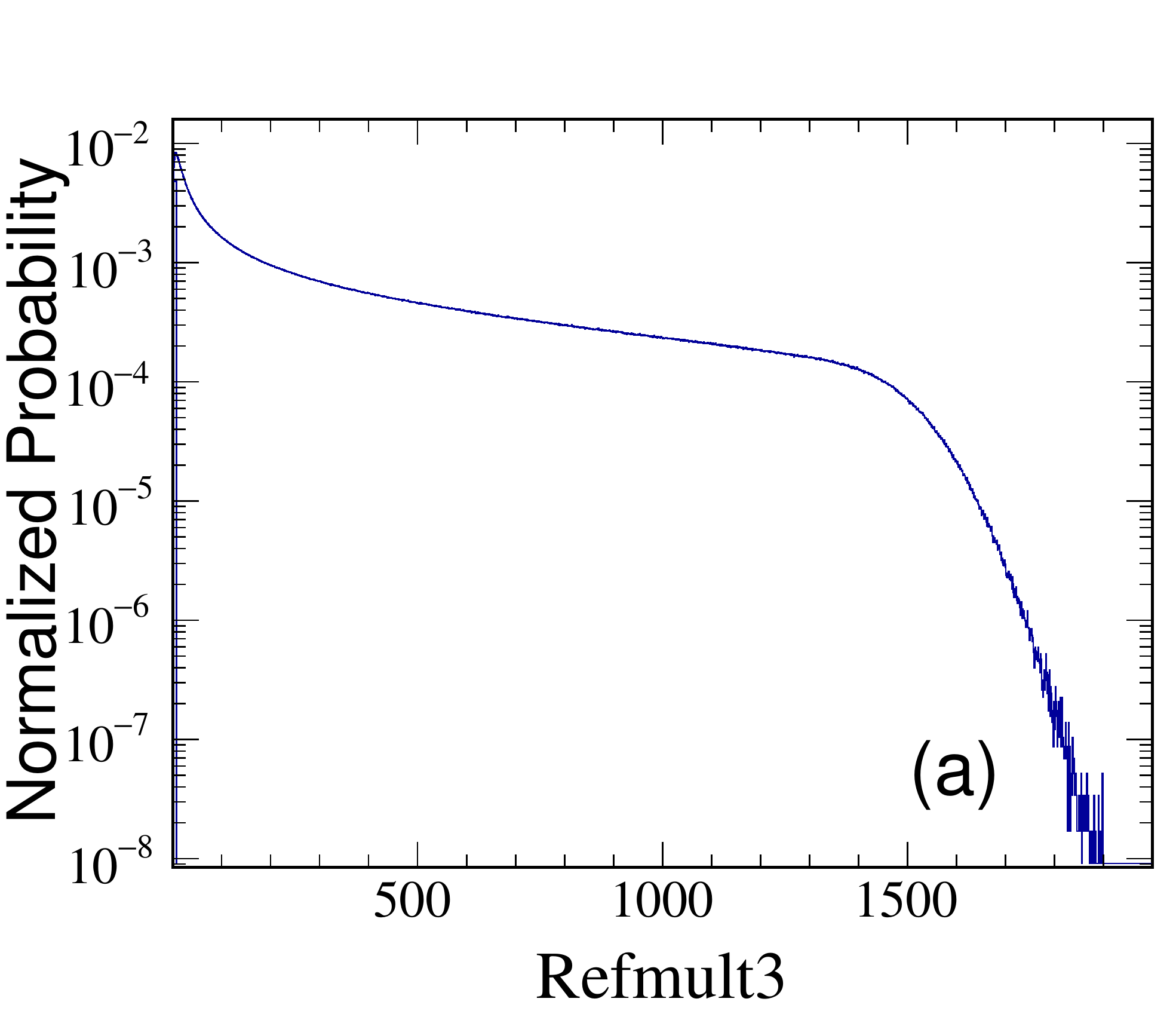}
\includegraphics[width=2.1in]{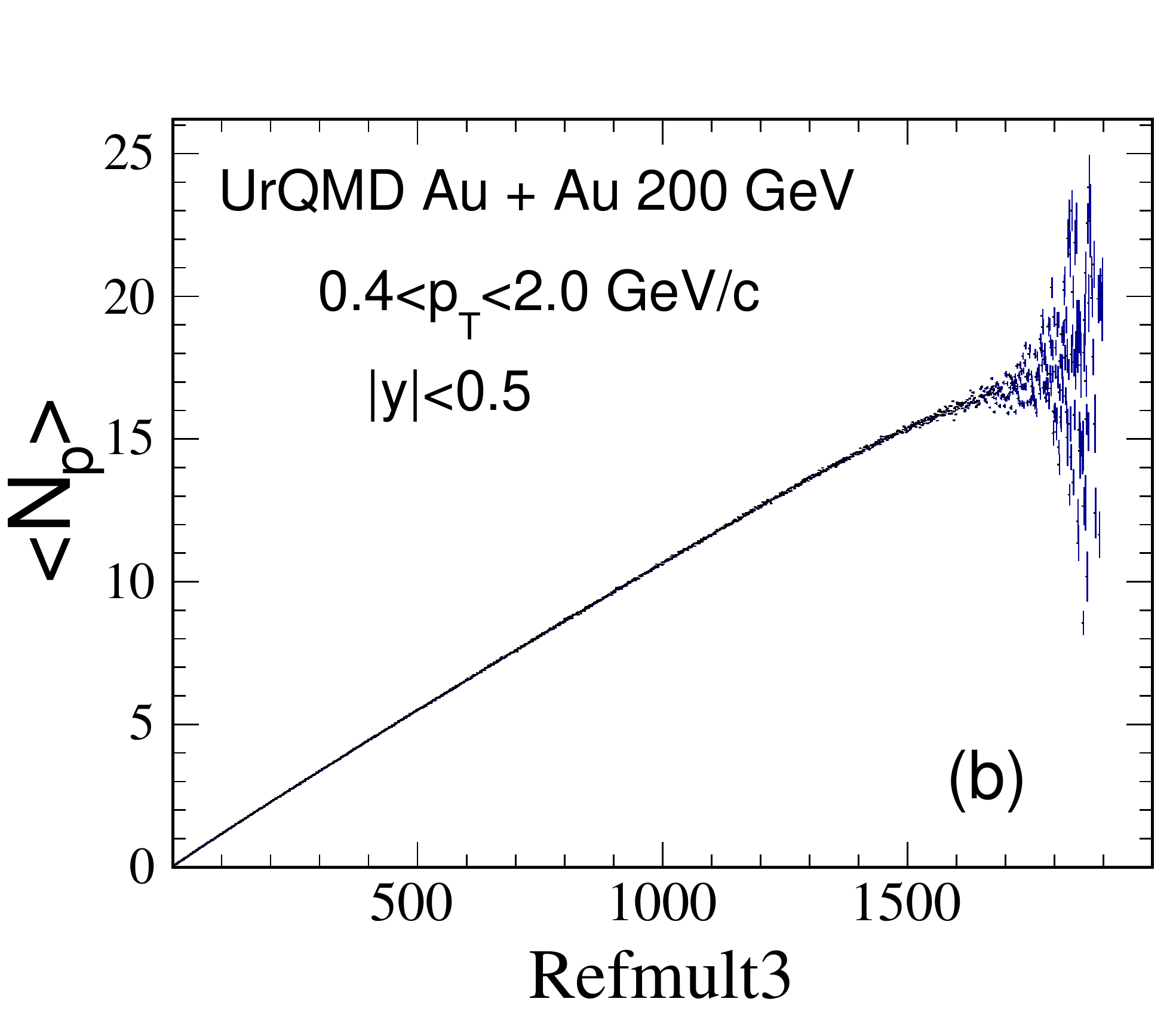}
\includegraphics[width=2.1in]{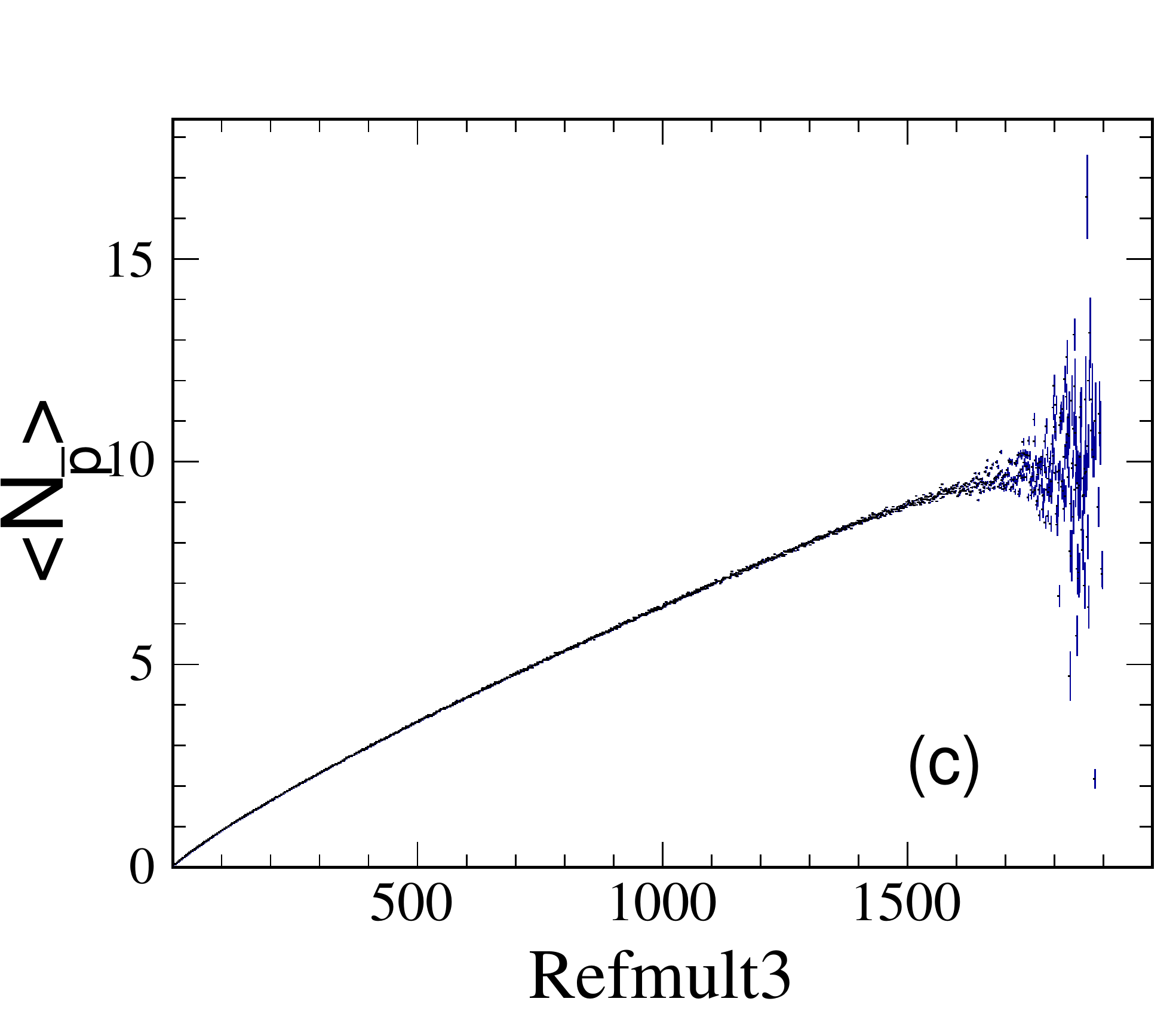}
\figcaption{\label{UrQMD-refmult3} (Color Online) The left panel shows normalized probability distribution of Refmult3 in UrQMD model in Au + Au collisions at $\sqrt{s_{NN}} = $ 200 GeV. The middle and right panels show  $\left<N_{p}\right>$  and $\left<N_{\bar{p}}\right>$ as a function of Refmult3, respectively. }
\end{center}
\end{figure*}

It is known that if the statistics are sufficient for validation of CLT~\cite{CLT}, then the experimentally measured results should always be consistent with the true value within 1$\sigma$, 2$\sigma$ and 3$\sigma$ of uncertainties with probability of 68.3$\%$, 95.5$\%$ and 99.7$\%$, respectively. For most of the measurements, such as the mean of the net-proton number, the required statistics for CLT are easily achieved in experiments. In this case, within the uncertainties, the mean of the measured observable needs to be a real value and independent of the statistics. i.e.,
\begin{equation}\label{mean}
\left<X\right>_{n_{1}} = \left<X\right>_{n_{2}} = \left<X\right>_{n_{3}},
\end{equation}
where $X$ is the measured observable. The subscripts $n_1$, $n_2$ and $n_3$ denote different statistics. As an example, Fig.~\ref{mean-clt}(a) shows the simulated $\left<Mean\right>$ of the net-proton number as a function of the statistics. The BD is applied in simulations with parameters: $\mu_{p} = 12.66, \mu_{\bar{p}} = 7.5, \varepsilon_{p} = \varepsilon_{\bar{p}} = 0.97$. $\mu_{p}$ and $\mu_{\bar{p}}$ are mean values of proton and anti-proton in $0-10\%$ centrality with $0.4<p_{T}<2.0$ GeV/c and $|y|<0.5$. For each data point, we randomly and independently generate 50 sub-samples with the fixed statistics to calculate $\left<Mean\right>$ as:
\begin{equation}\label{mean-simulation}
\left<Mean\right> = \frac{\sum_{i=1}^{i=N}(Mean)_i}{N},
\end{equation}
where $(Mean)_i$ is the averaged number of net-proton in $i_{th}$ sub-sample. There are two methods to estimate  error($\left<Mean\right>$).  One is obtained by the formula of the error propagation:
\begin{equation}\label{mean-error}
error(\left<Mean\right>) = \frac{\sqrt{\sum_{i=1}^{i=N}error(Mean)_i^{2}}}{N}.
\end{equation}
We can also first calculate the width of $\left<Mean\right>$ based on these $N$ results.
Then error($\left<Mean\right>$) is the width of $\left<Mean\right>$ divided by $\sqrt{N}$.
A Good agreement is obtained based on these two methods. In this paper, the formula of error propagation is used to measure the statistical uncertainty.

The simulated statistics in Fig.~\ref{mean-clt}(a) are from 50 to 5,000 events. The black dashed line is the theoretical expectation. For mean of the net-proton number analysis,  Fig.~\ref{mean-clt}(a) clearly shows that 50 events are  sufficient for the validation of the CLT. That is why we do not require to check whether the statistics are sufficient for most of the observables.

However, it is a challenge for analysis of the high-order cumulants which are up to the fifth order, sixth order, or even the eighth order~\cite{statistics-chenlz-v1, statistics-chenlz-v2, statistics-chenlz-v3}. When the CBWC method is applied in cumulants calculations, the statistics in each $N_{ch}$ are significantly limited in $0-10\%$ centrality even with a few hundred million MB events. As a crude estimation,  supposing there are 1000 $N_{ch}$ bins in 0-10$\%$ centrality, the averaged events  are only approximately 10,000 in each $N_{ch}$ with 100M MB events.
 If 10, 000 events are not sufficient for CLT in $C_6/C_2$ calculations, the value obtained by the CBWC method is not reliable in 0-10\% centrality with 100M MB events.

By using the same simulated parameters as Fig.~\ref{mean-clt}(a), while the simulated sub-sample $N$ is significantly larger than 50,  the statistics dependence of $\left<C_6/C_2\right>$ from 10,000 to 3M events are demonstrated in Fig.~\ref{mean-clt}(b). Below 0.1M events in each $N_{ch}$, $\left<C_6/C_2\right>$ is systematically smaller than 0 and theoretical BD expectations. Up to 0.5M events in each $N_{ch}$, $\left<C_6/C_2\right>$ is consistent with theoretical expectations within statistical uncertainties.

 As we mentioned,
 the required statistics are also related to the detail shape
and width of the net-proton multiplicity distributions~\cite{statistics-chenlz-v1, statistics-chenlz-v2, statistics-chenlz-v3}. The behavior of the statistical dependency in different centralities requires a careful case-by-case study.
 To directly compare, we study the statistics dependence of $C_6/C_2$ in BD with the same calculation method as RHIC/STAR.

\section{Simulation Setup}
\begin {figure*}
\begin{center}
\includegraphics[width=2.85in]{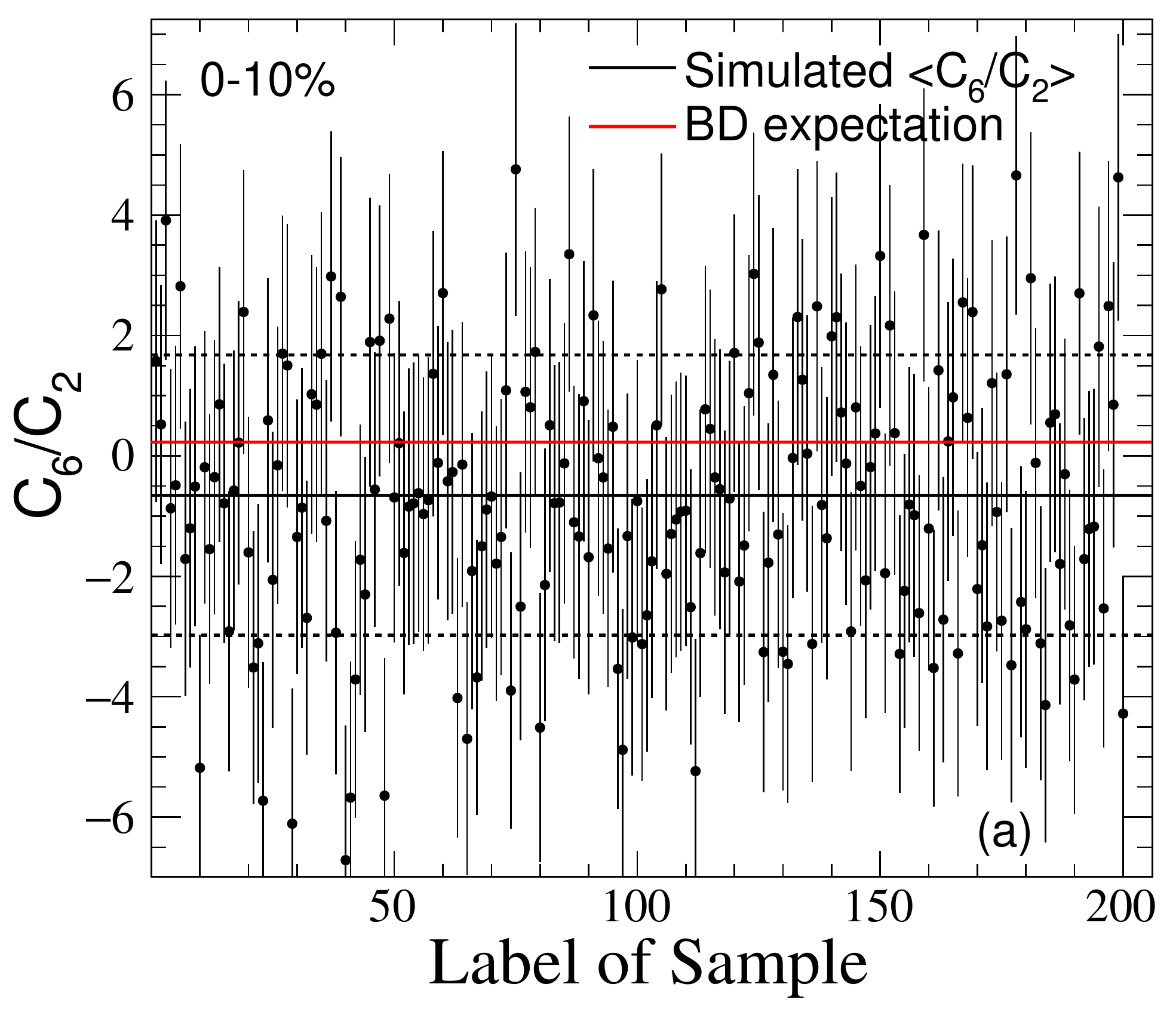}
\includegraphics[width=2.85in]{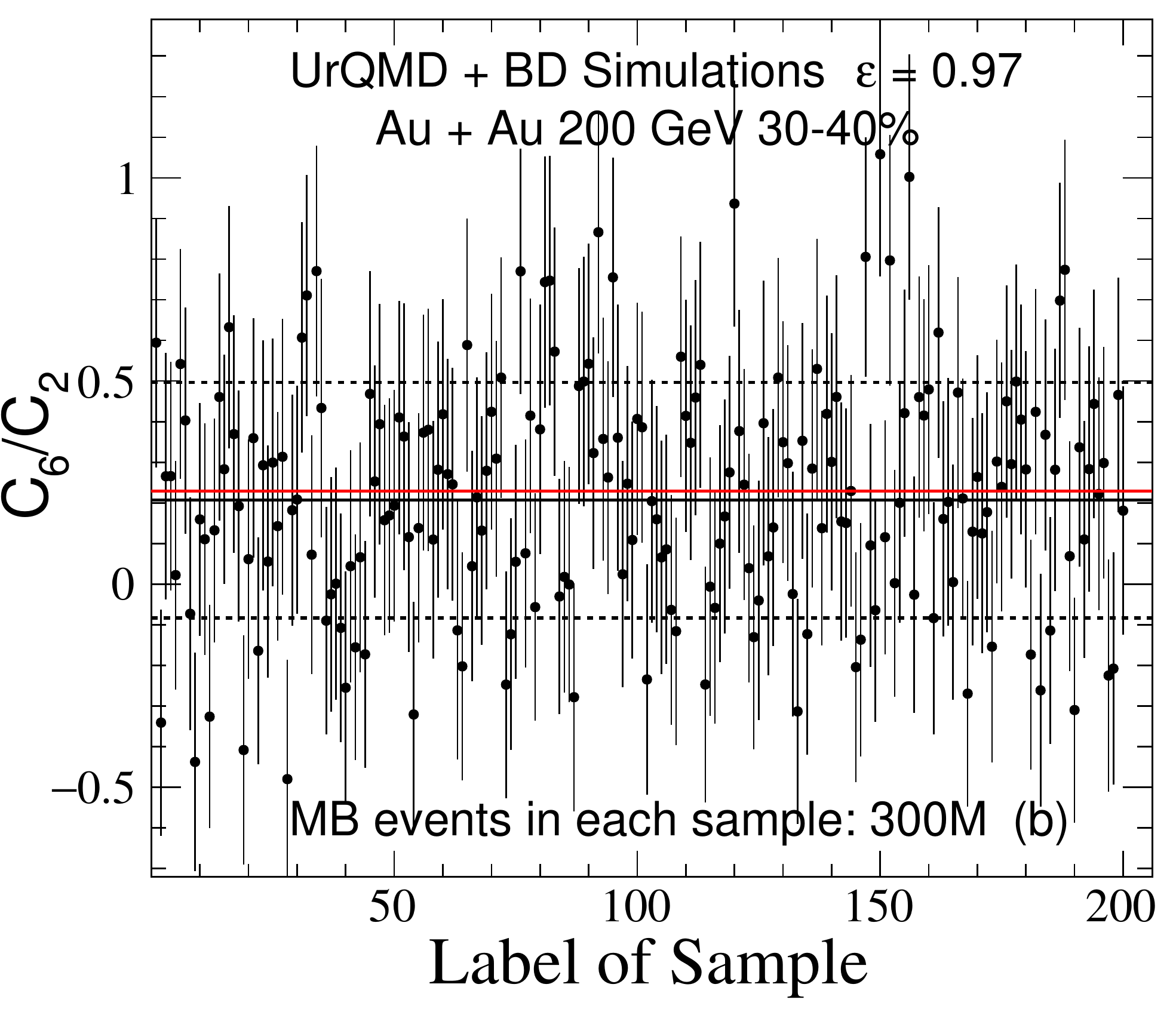}
\figcaption{\label{C6C2-bd-example} (Color Online)  $C_6/C_2$ and their statistical errors of 200 samples in 0-10\% and 30-40\% centrality, respectively. Each of $C_6/C_2$ is calculated by the CBWC method with 300M MB events.  The protons and anti-protons are independently and randomly generated from the BD with $\varepsilon = 0.97$. The input parameters for simulations are taken from UrQMD model at $\sqrt{s_{NN}} = $ 200 GeV in Au + Au collisions.}
\end{center}
\end{figure*}

To study the statistics dependence of   $C_6/C_2$ measurement based on the CBWC method, the input parameters for simulations are taken from the UrQMD model at $\sqrt{s_{NN}} = $ 200 GeV in Au + Au collisions~\cite{UrQMD-1, UrQMD-2}.

Fig.~\ref{UrQMD-refmult3}(a) shows the normalized probability distribution of Refmult3, which is used for the centrality selection.  To avoid the auto-correlation and improve the centrality resolution, Refmult3 is defined by using the number of charged $\pi$ and $K$ in the final state within  pseudo-rapidity $|\eta|<$1.0~\cite{cbwc-v3}.
Figs.~\ref{UrQMD-refmult3}(b) and (c) show the means of proton and anti-proton numbers,  $\left<N_{p}\right>$  and $\left<N_{\bar{p}}\right>$,  as a function of Refmult3, respectively. As a comparable study to  STAR measurements, the protons and anti-protons in the UrQMD model are carried out at mid-rapidity $(|y|<0.5)$ in the transverse momentum range $0.4<p_T<2.0$ GeV/$c$.  These values of $\left<N_{p}\right>$  and $\left<N_{\bar{p}}\right>$ are taken as the input parameters in the following simulations.

 For a straightforward comparison between the simulated results and theoretical expectations,  we assume that the numbers of protons and anti-protons independently follow BD with $\varepsilon = \varepsilon_{p} = \varepsilon_{\bar{p}} = 0.97 $ in each Refmult3 bin.  Finally, the influence of statistics on $C_6/C_2$ with the whole interval of $\varepsilon$, from 0.94 to 0.99, is studied. The required number of MB events  is  300 million (M).
Under these assumptions,  the simulated procedures are  implemented as follows:

(a) Multiplying the normalized probability of Refmult3 by 300M.  The corresponding number of events in $j_{th}$ Refmult3 bin is $n_j$.

(b) According to $n_{j}$, the values of   $\left<N_{p}\right>$  and $\left<N_{\bar{p}}\right>$ in $j_{th}$ Refmult3 bin, the numbers of protons and anti-protons are independently and randomly generated event-by-event by BD in each Refmult3 bin.

(c) Calculating $\left(C_6/C_2\right)_{j}$  and its error in each Refmult3 bin. Its error is estimated by the delta  theorem method~\cite{delta-error}. Then the centrality dependence of  $C_6/C_2$ and error($C_6/C_2$)  in each centrality are obtained  based on the CBWC method.

(d) Repeating the procedures of (b) and (c) $N$ times. Here, $N$ should be large enough to reduce the uncertainty of averaged $C_6/C_2$,  $\left<C_6/C_2\right>$. We  set $N = 1000$ in the following.

(e) Calculating the centrality dependence of $\left<C_6/C_2\right>$ and error($\left<C_6/C_2\right>$). The formula of error propagation is used to estimate the statistical uncertainty.

To demonstrate more details about the simulations, Figs.~\ref{C6C2-bd-example}(a) and (b) show  200 results of 1000 samples  in 0-10\% and 30-40\% centrality, respectively.
To reduce the statistical error and get a more stable result,  the value of $\left<C_6/C_2\right>$, shown by the black solid line in each panel, is derived from the total 1000 samples. The black dashed lines are 1$\sigma$ limits of $C_6/C_2$.
Both of these two plots clearly demonstrate that $C_6/C_2$ randomly fluctuates around  $\left<C_6/C_2\right>$.   In  Fig.~\ref{C6C2-bd-example}(a), the probabilities of $C_6/C_2$ that lie outside 1$\sigma$ and 2$\sigma$ of $\left<C_6/C_2\right>$  are about 31.0\% and 5.0\%, respectively.  Meanwhile, 30.0\% and 4.0\% of the observations lie outside of the  1$\sigma$ and 2$\sigma$  standard deviations shown in    Fig.~\ref{C6C2-bd-example}(b). Theoretically, for $N$ independent observations $(x_{1}, x_{2}, \cdots, x_{N})$ with the same expectation, they are approximated with a normal distribution.  In this case, 31.7\% and 4.5\% of the observations are  outside  1$\sigma$ and 2$\sigma$ standard deviations of the expectation, respectively.  Consequently, our results are consistent with the theoretical studies.  It confirms the validity of our simulations.

\begin{figure*}[htp]
\begin{center}
\includegraphics[width=4.8in]{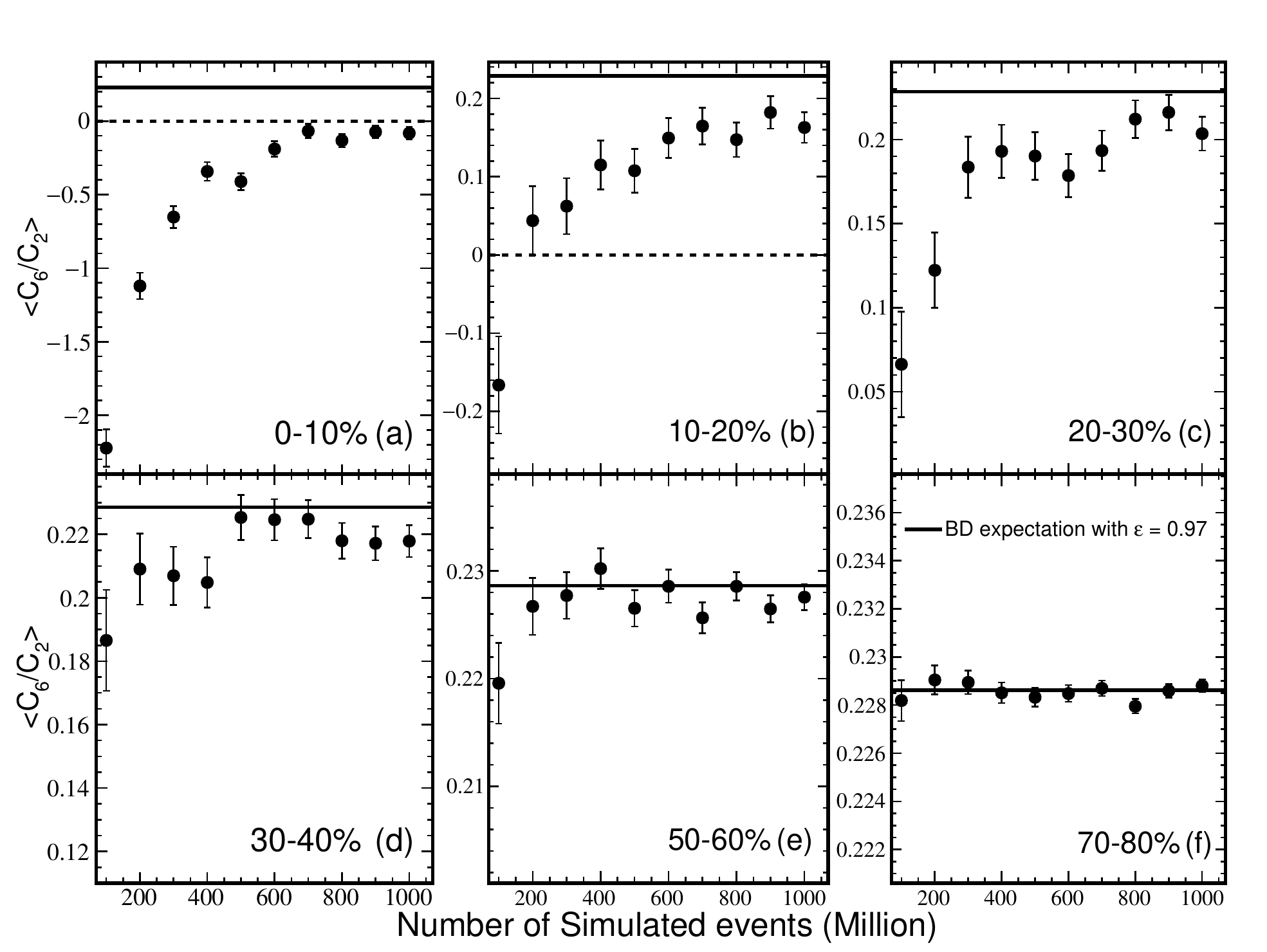}
\caption{\label{C6-bd-centrality}  Statistical dependence of  $\left<C_6/C_2\right>$ in BD with $\varepsilon = 0.97$.  The statistical error is calculated from formula of the error propagation. The black solid lines are the corresponding theoretical BD expectations derived from Eq.~(\ref{C6C2-bd}).}
\end{center}
\end{figure*}

The red solid line in each panel is the theoretical BD expectation. It is about  0.233 with $\varepsilon = 0.97$. In principle, if the statistics in each Refmult3 bin are sufficient, the final $\left<C_6/C_2\right>$ should be consistent with 0.233 in all centralities.  However,  Figs.~\ref{C6C2-bd-example}(a) and (b) show the disagreement of $C_6/C_2$ between the simulated $\left<C_6/C_2\right>$  and theoretical BD expectation. The difference of $C_6/C_2$ in 0-10\% centrality is more significant than that in 30-40\% centrality.  Comprehensive studies of the  statistics dependence of $C_6/C_2$ are shown in the following section.

\section{Statistics dependence of $\left<C_6/C_2\right>$ in BD}

As we mentioned, the experimental data collected for $C_6/C_2$ analysis include three different runs: 300M MB events, 420M MB events, and 110M from central trigger.  Here 100M events  from central trigger  can be reference to  1000M MB events in our simulations.
As a direct comparison, we can study the statistical dependence of $\left<C_6/C_2\right>$ with the magnitude of a few hundred million.

 By considering $\varepsilon=0.97$ to the simulations, Figs.~\ref{C6-bd-centrality}(a) to (f) show statistical dependence of $\left<C_6/C_2\right>$ in different centralities from 100M to 1000M MB events.    In 0-10\% centrality,  Fig.~\ref{C6-bd-centrality}(a) shows that all of the results are systematically smaller than the black solid line, which is the theoretical expectation.
Below 500M MB events, the values of $\left<C_6/C_2\right>$  first increase as the statistics increase and they are all negative.  From 600M to 1000M MB events,   $\left<C_6/C_2\right>$ is consistent with zero with 2$\sigma$ of the statistical uncertainty.  Consequently, with the current statistics at RHIC/STAR, $C_6/C_2$ may be under-estimated in 0-10\% centrality at $\sqrt{s_{NN}} = $ 200 GeV.

 In 10-20\% centrality,  Fig.~\ref{C6-bd-centrality}(b) shows that the values of $\left<C_6/C_2\right>$ are still systematically smaller than that in BD. However, it becomes positive above 200M MB events.  It suggests that the effect of the statistics becomes weaker.
 In  20-30\% and 30-40\% centralities, $\left<C_6/C_2\right>$ is only slightly smaller than that in BD, as shown in Figs.~\ref{C6-bd-centrality}(c) and (d). Finally, they are in agreement with each other in peripheral collisions shown in Figs.~\ref{C6-bd-centrality}(e) and (f).

Fig.~\ref{C6-bd-centrality} shows that the influence of statistics on $\left<C_6/C_2\right>$ is strongest in central collisions, comparing to the mid-central and peripheral collisions.
In 0-10\% centrality, $\left<C_6/C_2\right>$ is still smaller than the expectation even with 1000M MB events.
$\left<C_6/C_2\right>$ is still about 0 and we do not observe the convergence even with 1,000M MB events.
It is due to the wider distribution of net-proton number and the smaller statistics in each $N_{ch}$ in $0-10\%$ centrality. It can be further understood from Fig.~\ref{mean-clt}(b), in which the simulated parameters are also extracted from mean values of proton and anti-proton in $0-10\%$ centrality in the UrQMD model.  As the same estimation as in section 2, with 1,000M MB events, the averaged statistics in each $N_{ch}$ is around 0.1M, assuming there are 1,000 $N_{ch}$ bins in $0-10\%$ centrality.  Fig.~\ref{mean-clt}(b) also shows that $\left<C_6/C_2\right>$ is also about 0 with 0.1M events in each $N_{ch}$, which is consistent with that shown in Fig.~\ref{C6-bd-centrality}(a). In contrast, Fig.~\ref{mean-clt}(b) also shows that $\left<C_6/C_2\right>$ is consistent with the theoretical expectations above 0.5M events, which suggests that 5,000M MB events are required if the CBWC method in each $N_{ch}$ is applied. That is why the convergence is not observed in Fig.~\ref{C6-bd-centrality}(a).

 One possible alternative is to reduce the number of bins  when applying CBWC method, such as the centrality bin width of each $\delta 0.5\%$, $\delta 1.0\%$, and $\delta2.5\%$ and so on. It is beyond the scope of this paper to examine through the transport model or the experimental data directly.

\begin{center}
\includegraphics[width=3.0in]{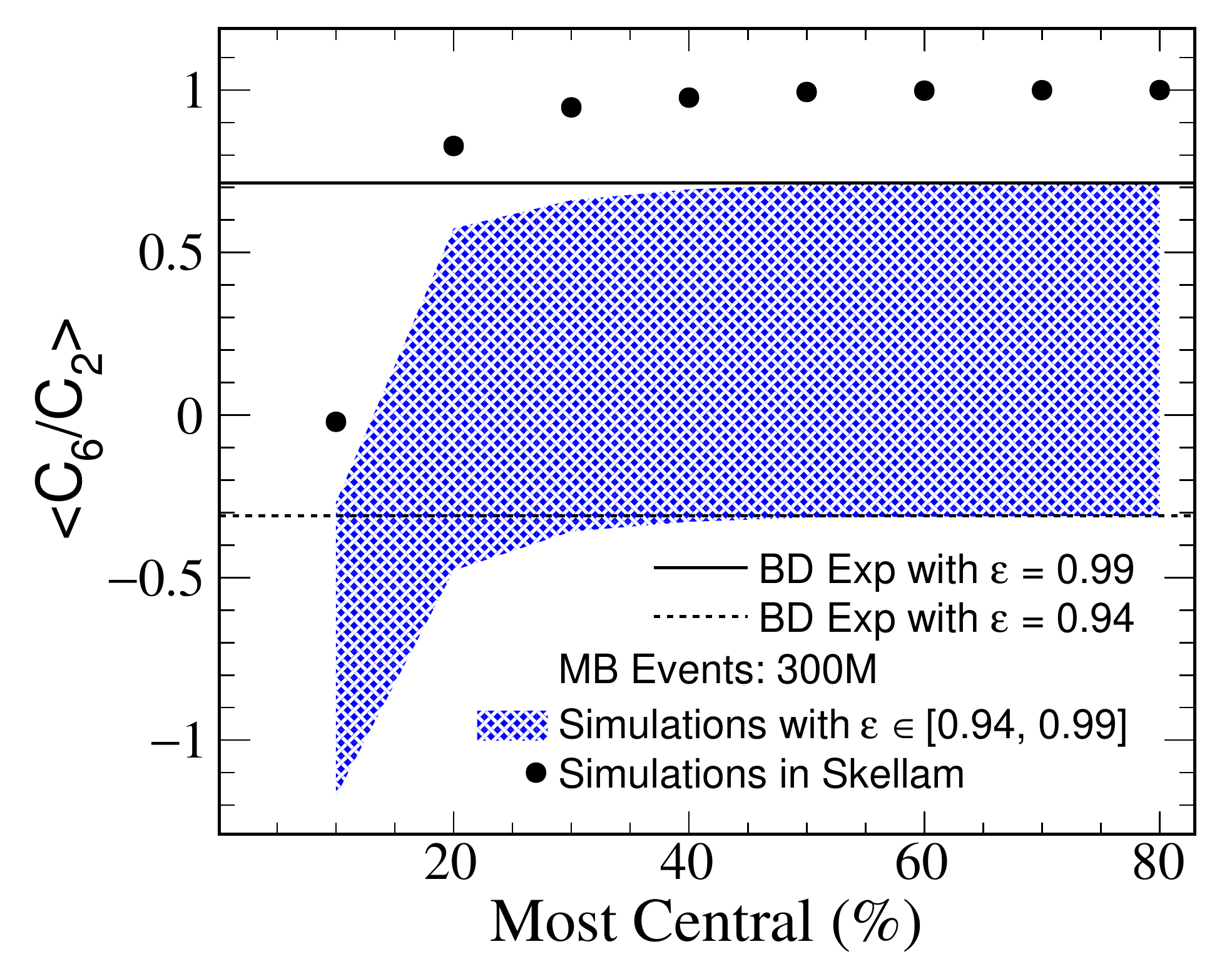}
\figcaption{\label{C6-bd-band} (Color Online)  Centrality dependence of $\left<C_6/C_2\right>$ in BD with $\varepsilon$ from 0.94 to 0.99 and in SD, respectively. Each $\left<C_6/C_2\right>$ is averaged from 10,000 randomly and independently simulated samples, with 300M MB events in each sample.}
\end{center}

Because the influence of the statistics  cannot be neglected, it is necessary to study $C_6/C_2$ in BD with the current statistics at RHIC/STAR.
With 300M MB events,  the blue filled band of Fig.~\ref{C6-bd-band} shows centrality dependence of $\left<C_6/C_2\right>$ with $\varepsilon$ from 0.94 to 0.99.   The solid and dashed lines are theoretically expected $C_6/C_2$ with $\varepsilon=0.99$ and 0.94, respectively.
 In  0-10\% and 10-20\% centralities, the simulated results are systematically smaller than those with the ideal theoretical expectations. Particularly,  there is a marked drop in $\left<C_6/C_2\right>$  in 0-10\% centrality.  With $\varepsilon=0.99$, the simulated $\left<C_6/C_2\right>$ is still negative, while the theoretical expectation is about 0.71.  Meanwhile, this  phenomenon is  also observed with the same simulation method in Skellam distribution (SD), shown by circular solid points in  Fig.~\ref{C6-bd-band}. The simulated $\left<C_6/C_2\right>$ is consistent with zero while it must be unity in SD.  It further confirm that $C_6/C_2$ can be under-estimated in 0-10\% centrality.  With insufficient statistics, $C_6/C_2$ can vary from a positive value to a negative one.

Together with the same statistics as RHIC/STAR, the negative  $C_6/C_2$ is more significant in BD.  The blue filled band shown in Fig.~\ref{C6-bd-band} almost covers the current preliminary measured  $C_6/C_2$ within experimental large uncertainties~\cite{STAR-C6C2-v1}.  It can better describe the experimental measured $C_6/C_2$, comparing to the SD baseline. Nonetheless,  it cannot completely reproduce the experimentally measured results.
First, the experimental data show $C_6/C_2$ decreases from peripheral to central collisions~\cite{STAR-C6C2-v1}. In our simulations, the variation range of $\left<C_6/C_2\right>$ is too large, although $\varepsilon$ is just from 0.94 to 0.99. It is difficult to study the feature of centrality dependence of $C_6/C_2$. Second, besides the statistics and BD,  $C_6/C_2$ can still be affected by many other complex contributions. We must correct those contributions well before relating the experimental measurements to our calculations. Third, we assume the proton and anti-proton are produced independently. However, they must have correlation, such as from resonance decay, which is not considered in this paper.
A direct comparison between the experimentally measured results and this baseline  requires further studies. In contrast, our studies also imply that we must exclude factors of non-phase transition related influences well before connecting the measured $C_6/C_2$ to the theoretical Lattice QCD calculations.

\section{Summary}

We studied the behavior of the net-proton $C_6/C_2$ in BD in Au + Au collisions at $\sqrt{s_{NN}} = $ 200 GeV.  By a reasonable approximation of $\varepsilon = \varepsilon_p = \varepsilon_{\bar{p}}$,  the net-proton $C_6/C_2$ is only dependent on $\varepsilon$ in BD.  With protons and anti-protons carried out in the mid-rapidity $|y|<0.5$ and $0.4<p_{T}<2.0$ GeV/$c$ in experiment, $\varepsilon_{p}$ and $\varepsilon_{\bar{p}}$ are close to each other and the values are between 0.94 and 0.99 in all centralities. In this region, $\varepsilon$ has a very significant effect on $C_6/C_2$. The values of $C_6/C_2$ decrease from 0.71 to -0.31 with $\varepsilon$ from 0.99 down to 0.94.  The negative
$C_6/C_2$ is observed in the  pure statistical BD. It suggests that the negative $C_6/C_2$ is not sufficient to be considered as an indication of a smooth crossover transition.

We also  simulated the statistics dependence of  $\left<C_6/C_2\right>$  based on CBWC calculation method.
Our method can compare the simulated results and theoretical expectations in a straightforward manner.  The significantly dropped signal is observed in 0-10\% centrality. With 300M MB events, the simulated $C_6/C_2$ is negative, while it is about 0.71 with $\varepsilon=0.99$ in BD.  This  phenomenon is  also observed with the same method of simulation in SD. The simulated $\left<C_6/C_2\right>$ is consistent with zero while the theoretical expectation is unity.  Moreover, with 1000M MB events, the simulated  $\left<C_6/C_2\right>$ is consistent with zero, while the theoretical expectation is about 0.233 with  $\varepsilon = 0.97$ in BD.
 Consequently, with the current statistics at RHIC/STAR, $C_6/C_2$ may be under-estimated in 0-10\% centrality at $\sqrt{s_{NN}} = $ 200 GeV.

With the experimental collected statistics, the negative  $C_6/C_2$ is more significant in BD.  It shows the values of $C_6/C_2$ have a broad range which can change from positive to negative. Comparing the baselines of $C_6/C_2$ in SD and BD,  the BD  could better describe the experimental measured $C_6/C_2$. Within large uncertainties at RHIC/STAR, the obtained baseline almost covers the current preliminary measured $C_6/C_2$.  Consequently, the exclusions of the non-phase transition related influences are required when using $C_6/C_2$ to study the chiral phase transition.
\end{multicols}

\vspace{-1mm}
\centerline{\rule{80mm}{0.1pt}}
\vspace{2mm}

\begin{multicols}{2}

\end{multicols}

\clearpage

\end{CJK*}
\end{document}